\documentclass[twocolumn,english]{revtex4}
\usepackage[T1]{fontenc}
\usepackage[latin9]{inputenc}
\usepackage{textcomp}
\usepackage{amsmath}
\usepackage{graphicx}

\makeatletter

\newcommand{\lyxmathsym}[1]{\ifmmode\begingroup\def\b@ld{bold}
  \text{\ifx\math@version\b@ld\bfseries\fi#1}\endgroup\else#1\fi}

\@ifundefined{textcolor}{}
{%
 \definecolor{BLACK}{gray}{0}
 \definecolor{WHITE}{gray}{1}
 \definecolor{RED}{rgb}{1,0,0}
 \definecolor{GREEN}{rgb}{0,1,0}
 \definecolor{BLUE}{rgb}{0,0,1}
 \definecolor{CYAN}{cmyk}{1,0,0,0}
 \definecolor{MAGENTA}{cmyk}{0,1,0,0}
 \definecolor{YELLOW}{cmyk}{0,0,1,0}
}

\makeatother

\usepackage{babel}
\begin{document}

\title{Co$_{2}$FeAl thin films grown on MgO substrates: Correlation between
static, dynamic and structural properties }

\author{M. Belmeguenai$^{1}$, H. Tuzcuoglu$^{1}$, M. S. Gabor$^{2}$, T.
Petrisor jr$^{2}$, C. Tiusan$^{2,3}$, D. Berling$^{4}$, F. Zighem$^{1}$,
T. Chauveau$^{1}$, S. M. Chérif$^{1}$ and P. Moch$^{1}$}

\affiliation{$^{1}$Laboratoire des Sciences des Procédés et des Matériaux, CNRS
- Université Paris 13, \textit{France}}

\affiliation{$^{2}$ Center for Superconductivity, Spintronics and Surface Science,
Technical University of Cluj-Napoca, Romania}

\affiliation{$^{3}$ Institut Jean Lamour, CNRS-Université de Nancy, France}

\affiliation{$^{4}$ Institut de Science des Matériaux de Mulhouse, CNRS-Université
de Haute-Alsace, France }
\begin{abstract}
Co$_{2}$FeAl (CFA) thin films with thickness varying from 10 nm to
115 nm have been deposited on MgO(001) substrates by magnetron sputtering
and then capped by Ta or Cr layer. X-rays diffraction (XRD) revealed
that the cubic $[001]$ CFA axis is normal to the substrate and that
all the CFA films exhibit full epitaxial growth. The chemical order
varies from the $B2$ phase to the $A2$ phase when decreasing the
thickness. Magneto-optical Kerr effect (MOKE) and vibrating sample
magnetometer measurements show that, depending on the field orientation,
one or two-step switchings occur. Moreover, the films present a quadratic
MOKE signal increasing with the CFA thickness, due to the increasing
chemical order. Ferromagnetic resonance, MOKE transverse bias initial
inverse susceptibility and torque (TBIIST) measurements reveal that
the in-plane anisotropy results from the superposition of a uniaxial
and of a fourfold symmetry term. The fourfold anisotropy is in accord
with the crystal structure of the samples and is correlated to the
biaxial strain and to the chemical order present in the films. In
addition, a large negative perpendicular uniaxial anisotropy is observed.
Frequency and angular dependences of the FMR linewidth show two magnon
scattering and mosaicity contributions, which depend on the CFA thickness.
A Gilbert damping coefficient as low as 0.0011 is found.
\end{abstract}


\maketitle

\section{Introduction }

The performances of spintronic devices depend on the spin polarization
of the current. Therefore, half metallic materials should be ideal
compounds as high spin polarized current sources to realize a very
large giant magnetoresistance, a low current density for current induced
magnetization reversal, and an efficient spin injection into semiconductors.
Theoretically, different kinds of materials, such as Fe$_{3}$O$_{4}$
{[}1, 2{]}, CrO$_{2}$ {[}3{]}, mixed valence perovskites {[}4{]}
and Heusler alloys {[}5, 6{]}, have been predicted to be half metals.
Moreover, the half metallic properties in these materials have been
experimentally demonstrated at low temperature. However, oxide half
metals have low Curie temperature ($T{}_{C}$) and therefore their
spin polarization is miserably low at room temperature. From this
point of view, Heusler alloys are promising materials for spintronics
applications, because a number of them have generally high $T{}_{C}$
{[}7{]} and therefore they may offer an alternative material choice
to obtain half metallicity even at room temperature. Furthermore,
their structural and electronic properties strongly depend on the
crystal structure. Recently, Heusler compounds have attracted considerable
experimental and theoretical interest, not only because of their half
metallic behaviour but also due to magnetic shape memory and inverse
magneto-caloric properties that they exhibit. One of the most important
Co-based full-Heusler alloys is Co$_{2}$FeAl (CFA). It has a high
$T{}_{C}$ ($T{}_{C}=1000$ K) {[}7{]} and, therefore, it is promising
for practical applications. Indeed, it can provide giant tunnelling
magnetoresistance ($360\%$ at RT) {[}8,9{]} when used as an electrode
in magnetic tunnel junctions. Furthermore, as we illustrate in our
present study, CFA presents the lowest magnetic damping parameter
among Heuslers. This low damping should provide significantly lower
current density required for spin-transfer torque (STT) switching,
particularly important in prospective STT devices. However, the integration
of CFA as a ferromagnetic electrode in spintronic devices requires
a good knowledge allowing for a precise control of its magnetic properties,
such as its saturation magnetization, its magnetic anisotropy, the
exchange stiffness parameter, the gyromagnetic factor and the damping
mechanisms monitoring its dynamic behaviour. In this paper we used
X-rays diffraction (XRD), ferromagnetic resonance in microstrip line
(MS-FMR) under in-plane and out of plane applied magnetic field, combined
with transverse biased initial inverse susceptibility and torque (TBIIST)
method, in order to perform a complete correlated analysis between
structural and magnetic properties of epitaxial Co$_{2}$FeAl thin
films grown on MgO(001) substrates. In addition, a detailed study
of the different relaxation mechanisms leading to the linewidth broadening
is presented.

\section{Samples preparation and experimental methods}

CFA films were grown on MgO(001) single-crystal substrates using a
magnetron sputtering system with a base pressure lower than $3\times10^{-9}$
Torr. Prior to the deposition of the CFA films, a 4 nm thick MgO buffer
layer was grown at room temperature (RT) by rf sputtering from a MgO
polycrystalline target under an Argon pressure of 15 mTorr. Next,
the CFA films, with variable thicknesses (10 nm$\leq d\leq115$ nm),
were deposited at RT by dc sputtering under an Argon pressure of 1
mTorr, at a rate of 0.1 nm/s. Finally, the CFA films were capped with
a MgO(4nm)/Cr(10nm) or with a MgO(4nm)/Ta(10nm) bilayer. After the
growth of the stack, the structures were ex-situ annealed at 600$^{o}$C
during 15 minutes in vacuum (pressure lower than $3\times10^{-8}$
Torr). The structural properties of the samples have been characterized
by XRD using a four-circle diffractometer. Their magnetic dynamic
properties have been studied by microstrip ferromagnetic resonance
(MS-FMR). 

\begin{figure}
\includegraphics[bb=20bp 190bp 290bp 595bp,clip,width=8.5cm]{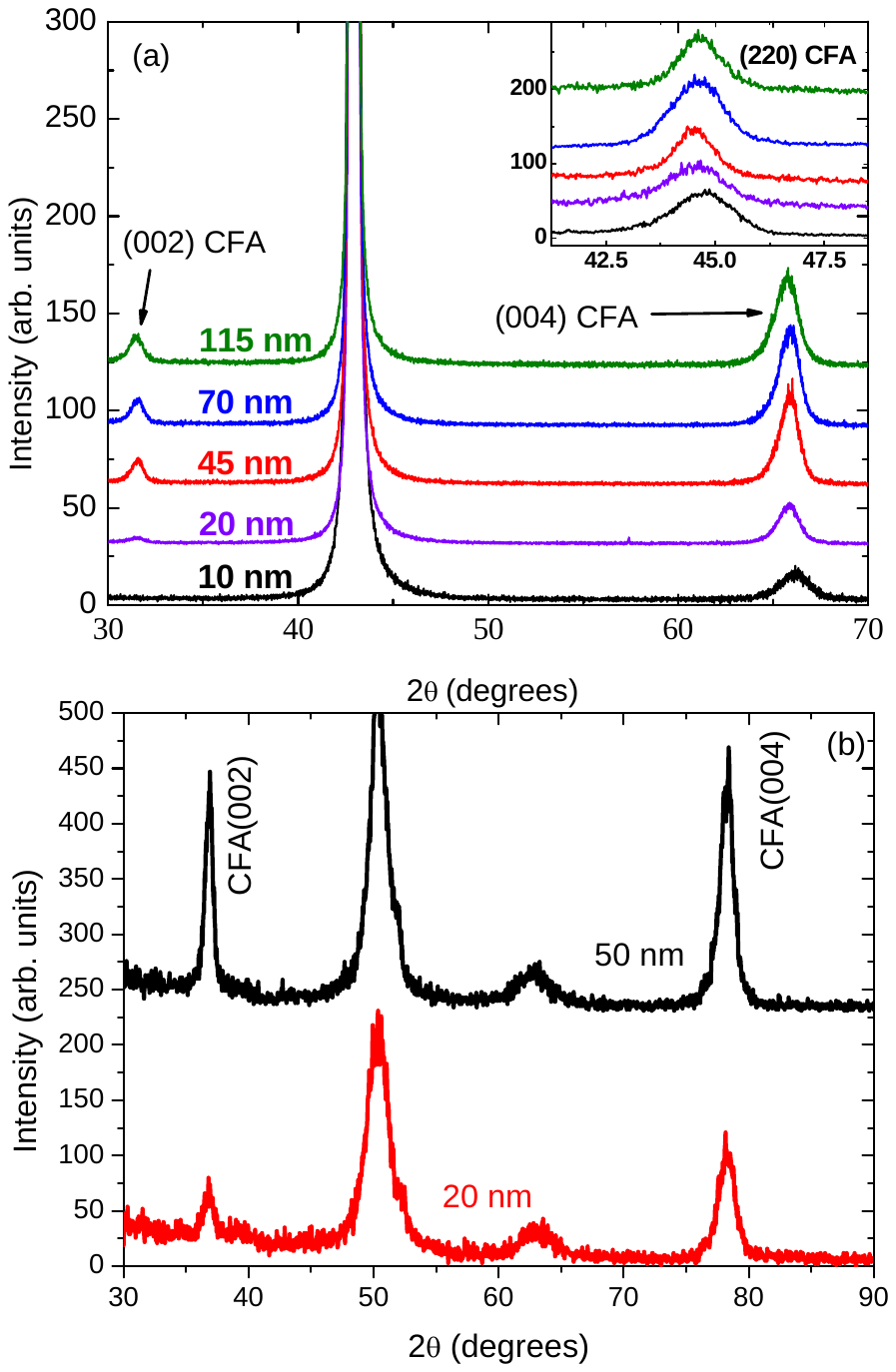}
\caption{(Colour online) (a) X-ray $2\theta-\omega$ (out-of-plane) diffraction
pattern using (Cu X-rays source) for the Cr-capped and (b) $\theta-2\theta$
pattern (Co X-rays source) for the Ta-capped Co$_{2}$FeAl of different
thicknesses. The inset shows selected area in plane diffraction patterns
around (220) Co$_{2}$FeAl reflection. }
\end{figure}

The MS-FMR characterization was done with the help of a field modulated
FMR setup using a vector network analyzer (VNA) operating in the 0.1-40
GHz frequency range. The sample (with the film side in direct contact)
is mounted on 0.5 mm microstrip line connected to the VNA and to a
lock-in amplifier to derive the field modulated measurements via a
Schottky detector. This setup is piloted via a Labview program providing
flexibility of a real time control of the magnetic field sweep direction,
step and rate, real time data acquisition and visualization. It allows
both frequency and field-sweeps measurements with magnetic fields
up to 20 kOe applied parallel or perpendicular to the sample plane.
In-plane angular dependence of resonance frequencies and fields are
used to measure anisotropies. The complete analysis of in-plane and
perpendicular field resonance spectra exhibiting uniform precession
and perpendicular standing spin wave (PSSW) modes leads to the determination
of most of the magnetic parameters: effective magnetization, gyromagnetic
factor, exchange stiffness constant and anisotropy terms. In addition,
the angular and the frequency dependences of the FMR linewidth are
used in order to identify the relaxation mechanisms responsible of
the line broadening and allow us for evaluating the parameters which
monitor the intrinsic damping (Gilbert constant) and the extrinsic
one (two magnon scattering, inhomogeneity, mosaïcity). 

Magnetization at saturation and hysteresis loops for each sample were
measured at room temperature using a vibrating sample magnetometer
(VSM) and a magneto-optical Kerr effect (MOKE) system. Transverse
biased initial inverse susceptibility and torque method (TBIIST) {[}10{]}
has been used to study the in-plane anisotropy for comparison with
MS-FMR. In this technique both a longitudinal magnetic sweep field
$\boldsymbol{H}_{L}$ (parallel to the incidence plane) and a static
transverse field $\boldsymbol{H}_{B}$ (perpendicular to the incidence
plane) are applied in the plane of the film and the longitudinal reduced
magnetization component $m_{L}$ is measured versus $H_{L}$ for various
directions of $\boldsymbol{H}_{L}$ with conventional magneto-optical
Kerr setup. From the measured hysteresis loops $m_{L}(H_{L})$ under
transverse biased field, the initial inverse susceptibility ($\chi^{-1}$)
and the field offset ($\delta H$) which are related to the second
and first angle-derivative of the magnetic anisotropy, respectively,
are derived. Fourier analysis of $\chi^{-1}$ and $\delta H$ versus
the applied field direction then easily resolves contributions to
the magnetic anisotropy of different orders and gives the precise
corresponding values of their amplitude and of their principal axes.

In order to obtain the desirable accuracy or even simply meaningful
results higher-order nonlinear in $m_{L}$ contributions (quadratic
or Voigt effect) as well as polar or other contributions to the Kerr
signal are carefully determined and corrected {[}10{]}. TBIIST method
surely does not have the same recognition than FMR techniques but
seems to be complementary, especially for samples with a weak magnetic
signal detectable with difficulty by FMR methods.

\section{Structural characterization}

Figure 1 shows the X-rays $2\theta-\omega$ diffraction patterns for
CFA of different thicknesses. These XRD patterns show that, in addition
to the feature arising from the (002) peak of the MgO substrate, the
Cr-capped samples (Fig. 1a: Cu X-rays source ($\lambda=0.15406$ nm))
exhibit only two peaks which are attributed to the (002) and (004)
diffraction lines of CFA. The Ta-capped films (Fig. 1b: Co-X-rays
source ($\lambda=1.7902$ $\mathrm{\AA}$)) show an additional peak
(around $2\theta=63\lyxmathsym{\textdegree}$) arising from the (002)
line issued from the Ta film. Pole figures (Fig. 2) allow to assert
an epitaxial growth of the CFA films according to the expected CFA(001){[}110{]}//MgO(001){[}100{]}
epitaxial relation. Using scans of various different orientations
we evaluated the out-of-plane ($a_{\perp}$) and the in-plane ($a_{\parallel}$)
lattice parameters (Fig. 3). A simple elastic model allowed us for
deriving the unstrained a0 cubic parameter as well as the in-plane
$\varepsilon_{\parallel}$ and the out-of- plane $\varepsilon_{\perp}$
strains:

\begin{figure}
\includegraphics[bb=20bp 340bp 290bp 595bp,clip,width=8.5cm]{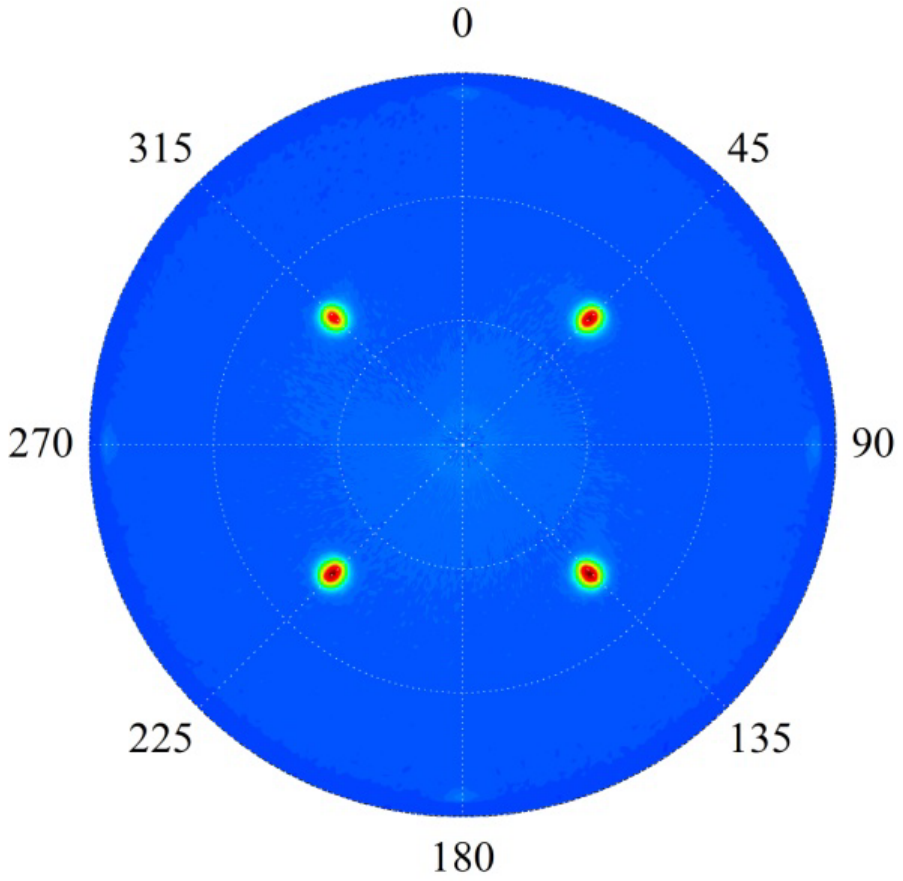}
\caption{(Colour online) Pole figures around the Co$_{2}$FeAl (022) type reflection,
for the 45 nm thick film, indicating the growth of Co$_{2}$FeAl on
MgO with the Co$_{2}$FeAl $(001)[110]\parallel$MgO$(001)[100]$
epitaxial relation. The 0 and 90 degrees axis of the graph correspond
to the MgO $[100]$ and {[}010{]} crystalline directions.}
\end{figure}

\begin{figure}
\includegraphics[bb=20bp 410bp 290bp 595bp,clip,width=8.5cm]{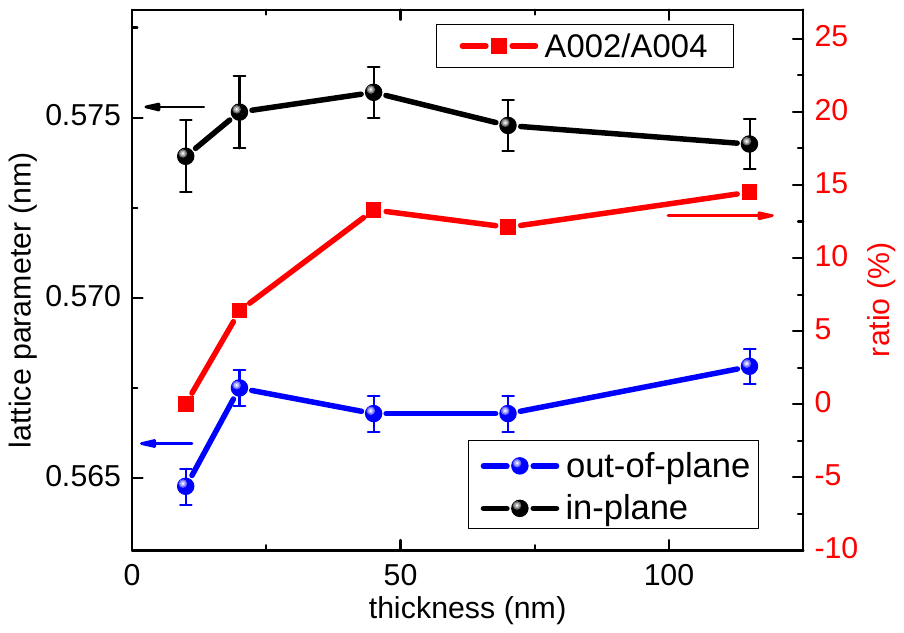}
\caption{(Colour online) Evolution of the out-of-plane and in-plane lattice
parameters and of the ratio of the integral intensities of the $(002)$
and $(004)$ Co$_{2}$FeAl peaks A(002)/A(004) with respect to the
Co$_{2}$FeAl films thickness.}
\end{figure}

\begin{multline}
a_{0}=\frac{\left(C_{11}a_{\perp}+2C_{12}a_{\parallel}\right)}{\left(C_{11}+2C_{12}\right)};\\
\varepsilon_{\parallel}=\frac{C_{11}}{\left(C_{11}+2C_{12}\right)}\frac{\left(a_{\parallel}-a_{\perp}\right)}{a_{0}};\\
\varepsilon_{\perp}=\frac{2C_{12}}{\left(C_{11}+2C_{12}\right)}\frac{\left(a_{\parallel}-a_{\perp}\right)}{a_{0}}
\end{multline}

where the values of the elastic constants $C_{11}=253$ GPa and $C_{12}=165$
GPa have been calculated previously {[}11{]}. Introducing the Poisson
coefficient $\nu=C_{12}/(C_{11}+C_{12})$ the above parameters write
as:

\begin{multline}
a_{0}=\frac{\left(\left(1-\nu\right)a_{\perp}+2a_{\parallel}\right)+2\nu a_{\parallel}}{\left(1+\nu\right)};\\
\varepsilon_{\perp}=\frac{\left(1-\nu\right)}{\left(1+\nu\right)}\frac{\left(a_{\parallel}-a_{\perp}\right)}{a_{0}};\\
\varepsilon_{\parallel}=-\frac{2\nu}{1+\nu}\frac{\left(a_{\parallel}-a_{\perp}\right)}{a_{0}}
\end{multline}

The cubic lattice constant $a_{0}$ does not depend upon the thickness,
except for the thinner 10 nm film (Fig. 4a), which shows a significant
reduction; its value,$0.5717\pm0.0005$ nm, is slightly smaller than
the reported one in the bulk compound with the L2$_{1}$ structure
(0.574 nm). The in-plane strain $\varepsilon_{\parallel}$ reveals
a tensile stress originating from the mismatch with the lattice of
the MgO substrate: however, its value does not exceed a few $^{\text{\textdegree}}$/$_{\text{\textdegree\textdegree}}$,
well below the Heussler/MgO mismatch, thus excluding an efficient
planar clamping. The strain $\varepsilon_{\parallel}$ decreases versus
the thickness, at least above 40 nm (Fig. 4b). 

Odd Miller indices (e.g.: $(111),(311),\lyxmathsym{\ldots}$) are
allowed for diffraction in the L2$_{1}$ phase {[}12{]}. In contrast,
they are forbidden in the B2 phase, which is characterized by a total
disorder between Al and Fe atoms but a regular occupation of the Co
sites. In the A2 phase the chemical disorder between Fe, Co and Al
sites is complete: $(hkl)$ diffraction is only allowed for even indices
subjected to $h+k+l=4n$. We do not observe $(111)$ or $(311)$ lines
and then conclude to the absence of the L2$_{1}$ phase in the studied
films. In contrast, a $(002)$ peak is observed, thus indicating that
the samples do not belong to the A2 phase. However, the ratio $I_{002}/I_{004}$
of the integrated intensities of the $(002)$ and of the $(004)$
peaks increases versus the film thickness (Fig. 3). This ratio is
proportional to $(1-2c)^{2}$, where $c$ is the chemical disorder.
Assuming that the thickest film belongs to the B2 phase ($c=0$) the
dependence of $c$ upon the film thickness is shown in figure 4c:
the A2 phase ($c=0.5$) is almost completely achieved for the 10 nm
thick sample. The reduction of $a_{0}$ in the thinner sample is probably
due to its previously noticed {[}13{]} smaller value in the A2 phase
compared to the B2 one. 

\begin{figure}
\includegraphics[bb=20bp 410bp 290bp 595bp,clip,width=8.5cm]{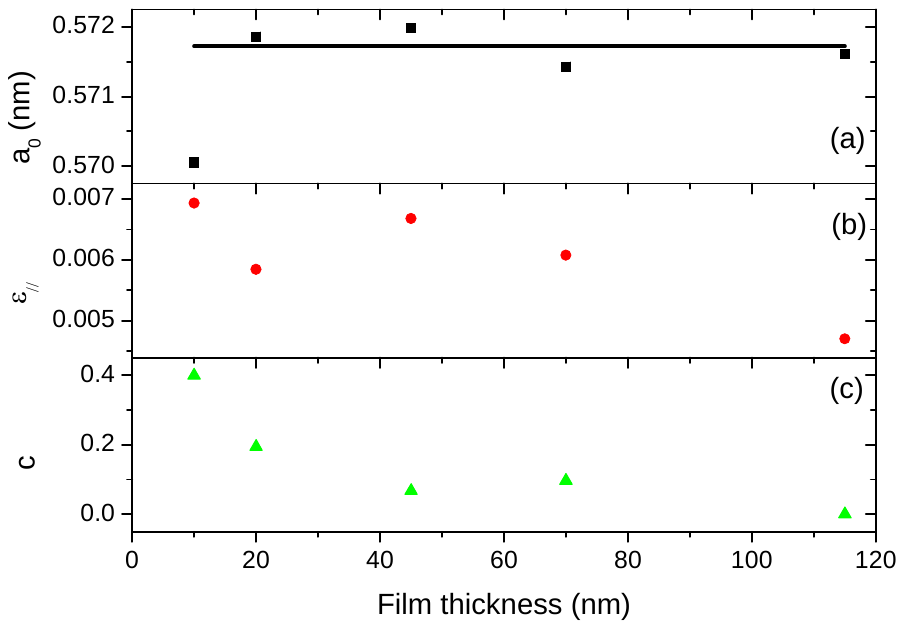}
\caption{(Colour online) Thickness dependence of (a) the lattice cubic parameter
$a_{0}$, the in-plane strain $\varepsilon_{\parallel}$ and (c) the
chemical order $c$ of Co$_{2}$FeAl thin films.}
\end{figure}

\section{Magnetic properties}

The experimental magnetic data have been analyzed considering a magnetic
energy density which, in addition to Zeeman, demagnetizing and exchange
terms, is characterized by the following effective anisotropy contribution
{[}14{]}:

\begin{multline}
E_{anis.}=-\frac{1}{2}(1+cos(2(\varphi_{M}-\varphi_{u}))K_{u}\sin^{2}\theta_{M}+\\
K_{\perp}\sin^{2}\theta_{M}-\frac{1}{8}(3+\cos4(\varphi_{M}-\varphi_{4}))K_{4}\sin^{4}\theta_{M}
\end{multline}

In the above expression, $\theta_{M}$ and $\varphi_{M}$ respectively
represent the out-of-plane and the in-plane (referring to the substrate
edges) angles defining the direction of the magnetization $M_{S}$.
$\varphi_{u}$ and $\varphi_{4}$ define the angles between an easy
uniaxial planar axis or an easy planar fourfold axis, respectively,
with respect to this substrate edge. $K_{\mathit{u}}$, $K_{\mathit{4}}$
and $K_{\perp}$ are in-plane uniaxial, fourfold and out-of-plane
uniaxial anisotropy constants, respectively. We introduce the effective
magnetization $M_{eff}=H_{eff}/4\pi$ obtained by: 

\begin{equation}
4\pi M_{eff}=H_{eff}=4\pi M_{S}-\frac{2K_{\perp}}{M_{S}}=4\pi M_{S}-H_{\perp}
\end{equation}

As experimentally observed, the effective perpendicular anisotropy
term $K_{\perp}$ (and, consequently, the effective perpendicular
anisotropy field $H_{\perp}$ ), is thickness dependent: $K_{\perp}$
describes an effective perpendicular anisotropy term which writes
as: 
\begin{equation}
K_{\perp}=K_{\perp V}+\frac{2K_{\perp S}}{d}
\end{equation}
 where $K_{\perp S}$ refers to the perpendicular anisotropy term
of the interfacial energy density. Finally we define $H{}_{\mathit{u}}=2K_{u}/M_{S}$
and $H_{4}=4K{}_{\mathit{4}}/M_{s}$ as the in-plane uniaxial and
the fourfold anisotropy fields respectively. The resonance expressions
for the frequency of the uniform and PSSW modes assuming in-plane
or perpendicular applied magnetic fields are given by equations (6)
and (7) respectively {[}14, 15{]}. 

\begin{multline}
F_{n.}=\frac{\gamma}{2\pi}(H\cos(\varphi_{H}-\varphi_{M})+\frac{2K_{4}}{M_{S}}\cos4(\varphi_{M}-\varphi_{4})\\
+\frac{2K_{u}}{M_{S}}\cos2(\varphi_{M}-\varphi_{u})+\frac{2A_{ex.}}{M_{S}}\left(\frac{n\pi}{d}\right)^{2})\times\\
(H\cos(\varphi_{H}-\varphi_{M})+4\pi M_{eff}+\frac{K_{4}}{2M_{S}}(3+\cos4(\varphi_{M}-\varphi_{4}))\\
+\frac{K_{u}}{M_{S}}(1+\cos2(\varphi_{M}-\varphi_{u}))+\frac{2A_{ex.}}{M_{S}}(\frac{n\pi}{d})^{2})
\end{multline}

\begin{multline}
F_{\perp.}=\frac{\gamma}{2\pi}(H-4\pi M_{eff}+\frac{2A_{ex.}}{M_{S}}\left(\frac{n\pi}{d}\right)^{2})\times
\end{multline}

In the above expressions \textit{$\gamma/2\pi=g\times1.397\times10^{6}$}
s$^{-1}$.Oe$^{-1}$ is the gyromagnetic factor, $n$ is the index
of the PSSW and $A_{ex}$ is the exchange stiffness constant.

The experimental results concerning the measured peak-to-peak FMR
linewidths $\Delta H^{PP}$ are analyzed in this work taking account
of both intrinsic and extrinsic mechanisms. Therefore, in the most
FMR experiments, the observed magnetic field linewidth ($\Delta H^{PP}$
) is usually analyzed by considering four different contributions
as given by equation (8) {[}16-21{]}. 
\begin{equation}
\Delta H^{PP}=\Delta H^{Gi}+(\Delta H^{mos}+\Delta H^{inh}+\Delta H^{2mag})
\end{equation}
When the applied field and the magnetization are parallel, the intrinsic
contribution is not angular dependent; it derives from the Gilbert
damping and is given by:

\begin{equation}
\Delta H^{Gi}=\frac{2}{\sqrt{3}}\frac{\alpha}{\gamma}2\pi f
\end{equation}

(9) where $f$ is the driven frequency and $\alpha$is the Gilbert
coefficient. 

The relevant mechanisms {[}16{]} describing the extrinsic contributions
are: 

1- Mosaicity: the orientation spread of the crystallites contributes
to the linewidth. Its contribution is given by: 

\begin{equation}
\Delta H^{mos}=\left|\frac{\partial H_{res}}{\partial\varphi_{H}}\Delta\varphi_{H}\right|=\left|\frac{\partial H}{\partial\varphi_{H}}\Delta\varphi_{H}\right|_{res}
\end{equation}

Where $\Delta\varphi{}_{H}$ is the average spread of the easy axis
anisotropy direction in the film plane. It is worth to mention that
for frequency dependent measurements along the easy and hard axes
the partial derivatives are zero and thus the mosaicity contribution
vanishes. The suffix ``res'' indicates that equation (10) should
be evaluated at the resonance. Therefore, using equation (6) for uniform
mode ($n=0$), the expression of $\frac{\partial H}{\partial\varphi_{H}}$
is found and then calculated using the corresponding value of $H$
and $\varphi_{M}$ at the resonance.

2- Inhomogeneous residual linewidth $\Delta H^{inh}$ present at zero
frequency. This contribution is frequency and angle independent inhomogeneity
related to various local fluctuations such as the value of the film
thickness. 

3- Two magnon scattering contribution to the linewidth. This contribution
is given by {[}22-24{]}: 

\begin{multline}
\Delta H^{2mag}=\Gamma_{0}+\Gamma_{2}\cos2(\varphi_{H}-\varphi_{2})+\\
\Gamma_{4}\cos4(\varphi_{H}-\varphi_{4})\arcsin\left(\frac{f}{\sqrt{f^{2}+f_{0}^{2}}+f_{0}}\right)
\end{multline}
 with: $f_{0}=\gamma M_{eff}$. The expected fourfold symmetry induces
the $\Gamma_{0}$ and $\Gamma_{4}$ coefficients; the coefficient
$\Gamma_{2}$ is phenomenogically introduced. 

The analysis of the variation of the resonance linewidth $\Delta H^{PP}$
versus the frequency and the in-plane field orientation allows for
evaluating $\alpha$, $\Delta\varphi_{H}$, $\Delta H^{inh}$, $\Gamma_{0}$,
$\Gamma_{2}$ (and $\varphi_{2}$) and $\Gamma_{4}$ (and $\varphi_{4}$
which, from symmetry considerations, is expected to have a $0\lyxmathsym{\textdegree}$
or $45\lyxmathsym{\textdegree}$ value, depending upon the chosen
sign of $\Gamma_{4}$).

\begin{figure}
\includegraphics[bb=20bp 0bp 290bp 595bp,clip,width=8.5cm]{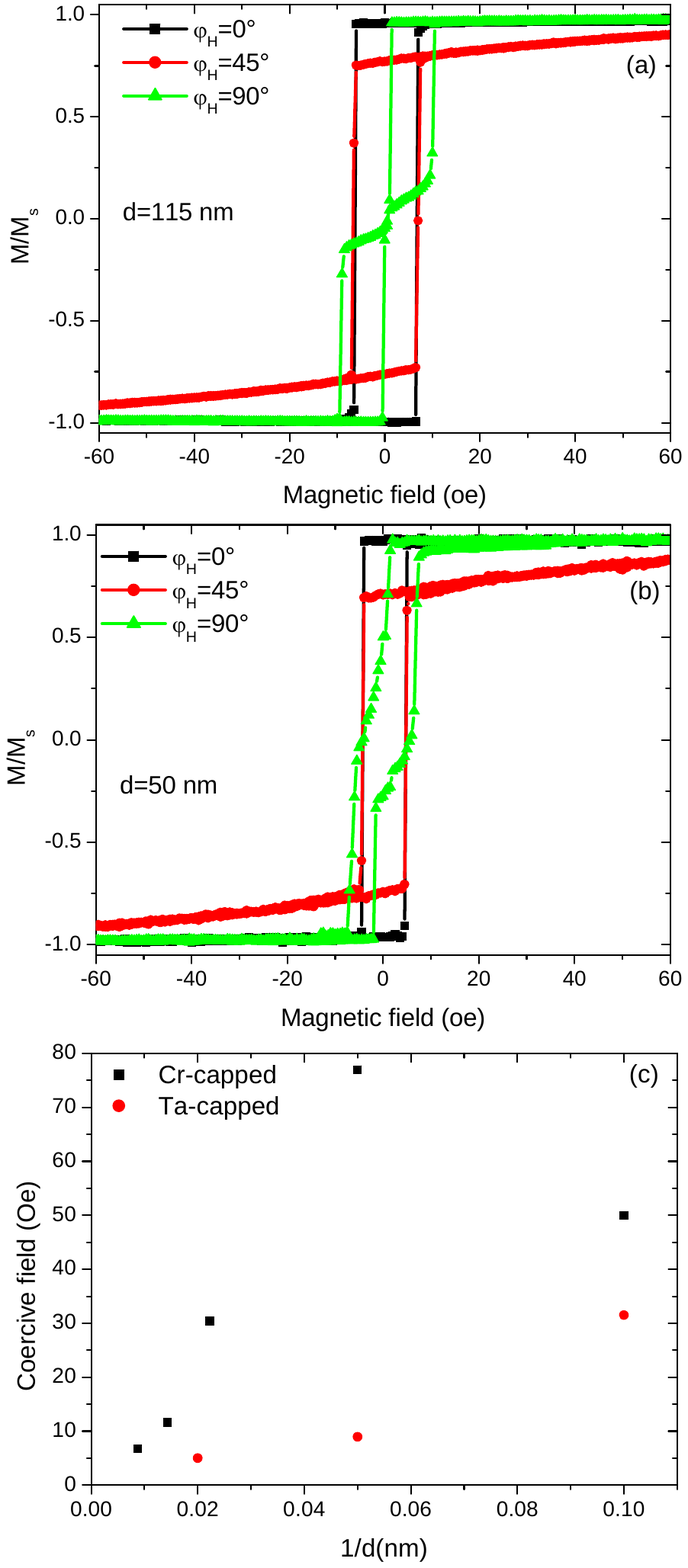}
\caption{(Colour online) MOKE hysteresis loops of the (a) 115 nm Cr-capped
and (b) 50 nm Ta-capped Co$_{2}$FeAl thin films. The magnetic field
is applied parallel to the film surface, at various angles ($\varphi_{H}$)
with a respect to edges of the MgO substrate ($[100]$ or $[010]$).
(c) Thickness dependence of the coercive field, deduced from hysteresis
loops along the easy axis, of Co$_{2}$FeAl Cr- and Ta-capped thin
films.}
\end{figure}

\subsection{Static properties }

The magnetization at saturation measured by VSM, averaged upon all
the samples has been found to be $M_{S}=1000\pm50$ emu/cm$^{3}$,
thus providing a magnetic moment of 5.05\textpm{}0.25 Bohr magneton
($\mu_{B}$) per unit formula, in agreement with the previously published
values for the B2 phase {[}7{]}. For all the studied films the hysteresis
loops were obtained by VSM and MOKE with an in-plane magnetic field
applied along various orientations. Figure 5 shows representative
behaviors of different CFA films. The observed shape mainly depends
on the field orientation, in agreement with the expected characteristics
of the magnetic anisotropy. As confirmed below, in all the studied
samples this anisotropy consists into the superposition of a fourfold
and of a uniaxial term showing parallel easy axes: this common axis
coincides with one of the substrate edges and, consequently, with
one of the $<110>$ crystallographic directions of the CFA phase.
It results that if an orientation (say $\varphi_{H}=0$ related to
$[110]$) is the easiest, the perpendicular direction (($\varphi_{H}=90^{\circ}$
) related to $[1\overline{1}0]$) is less easy. A similar situation
was studied and interpreted previously {[}25{]}: it is expected to
provide square hysteresis loops for $\varphi_{H}=0^{\circ}$ , as
evidenced in figure 5, while in contrast, for $\varphi_{H}=90^{\circ}$
, it leads to a two steps reversal, as can be seen in figure 5. The
intermediate step leads to a magnetization nearly perpendicular to
the applied field. For all the studied films a two steps loop is observed
for $\varphi_{H}$ ranging in the $\{55-130\lyxmathsym{\textdegree}\}$
interval. In figure 5c the deduced coercive fields ($H_{C}$) from
hysteresis loops along the easy direction ($\varphi_{H}=0^{\circ}$)
are compared for different thicknesses (10, 20, 45, 50, 70, and 115
nm). For both Cr-capped and Ta-capped films HC increases linearly
with the inverse of the film thickness. The Cr-capped samples present
higher coercive fields due to the different interface quality. 

\begin{figure}
\includegraphics[bb=20bp 210bp 290bp 595bp,clip,width=8.5cm]{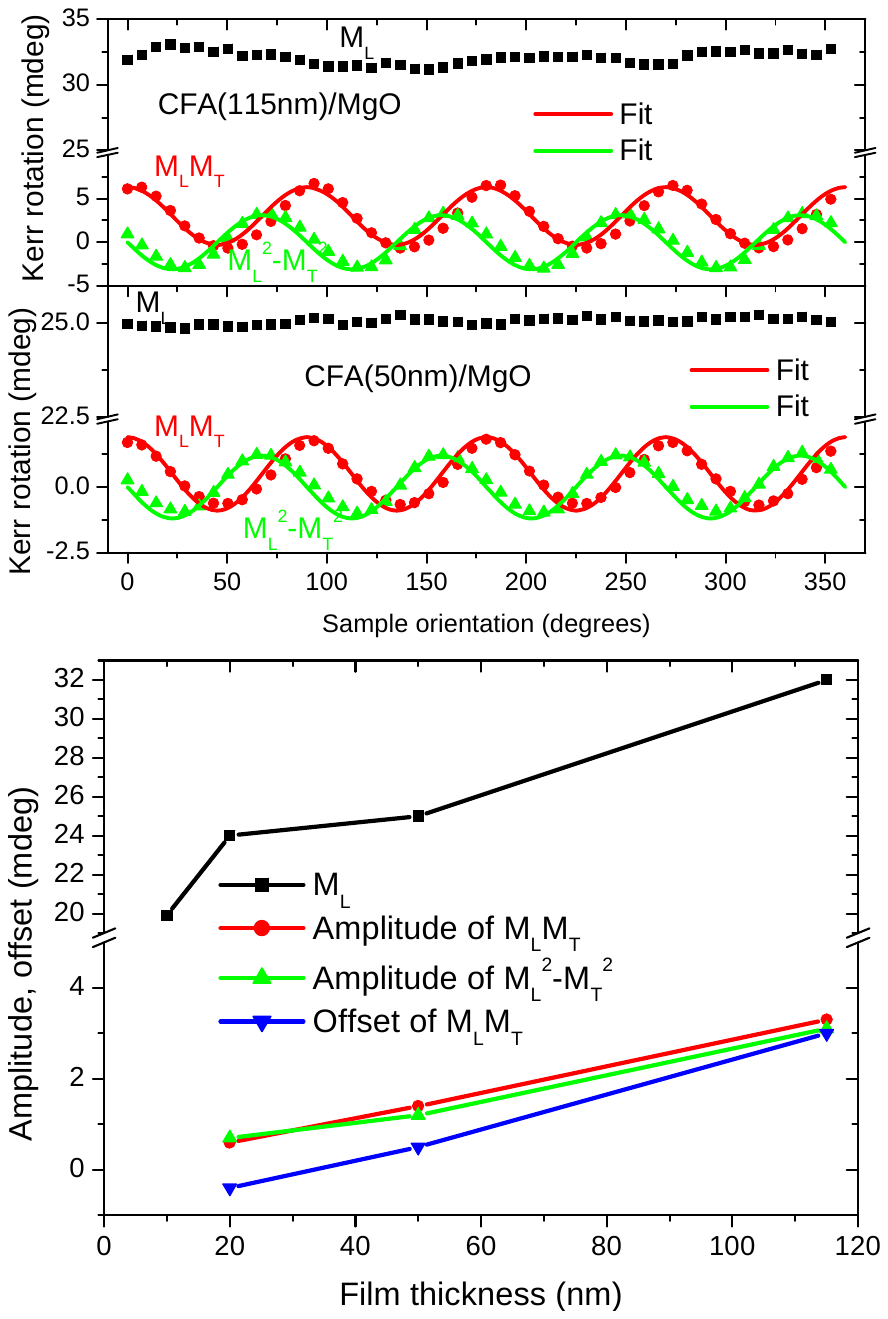}
\caption{(Colour online) (a) Separated quadratic MOKE contributions as a function
of the sample orientation at $46\lyxmathsym{\textdegree}$ incidence.
The fits are obtained using equation (12). (b) The $M_{L}$ contribution
(at angle of incidence of $46\lyxmathsym{\textdegree}$), the amplitudes
and offset of the $M_{L}M_{T}$ contribution and the amplitude of
the ($M_{L}^{2}-M_{T}^{2}$ ) as a function of the Co$_{2}$FeAl thickness.}
\end{figure}

One can also observe that MOKE hysteresis loops are not strictly centrosymmetrical
(see for example Fig. 5b for $\varphi_{H}=90^{\circ}$) indicating
the superposition of symmetrical (even function of applied sweep field
$H_{L}$) and anti-symmetrical (odd in $H_{L}$) components in the
Kerr signal. It has been shown and confirmed {[}26, 27{]} that, for
in-plane magnetized thin films, the antisymmetrical part observed
in the $m_{L}(H_{L})$ loops arises from the second order magneto-optical
effects quadratic in magnetization. Therefore, the present study was
not limited to the usual linear MOKE. We have also investigated this
quadratic contribution through the study of the Kerr signal dependence
upon the film orientation under a saturating in-plane field. Within
the cubic approximation for a (001) surface, the Kerr rotation angle
writes as {[}27{]}: 

\begin{multline}
\theta_{K}=a_{1}M_{L}+a_{2}(M_{L}^{2}-M_{T}^{2})\sin(4\psi)+\\
(b_{2}+2a_{2}\cos(4\psi))M_{L}M_{T}
\end{multline}

Where $M_{L}$ and $M_{T}$ stand for the longitudinal (i.e.: within
the incidence plane) and the transverse (i.e.: normal to the incidence
plane) component of the magnetization, respectively, and where $\psi$
is the angle of a cubic $<110>$ axis with the plane of incidence.
The first term describes the usual linear contribution while the following
ones correspond to the quadratic MOKE (QMOKE). The experimental study
was performed under an angle of incidence of $46\lyxmathsym{\textdegree}$
using a field magnitude large with respect to the anisotropy field.
The different contributions to the Kerr signal, as functions of the
film orientation $\psi$ are extracted by applying a rotating field
technique {[}10{]}. Representative results obtained with 115 Cr- and
50 nm Ta-capped films are shown in figure 6. Beside the longitudinal
component ($M_{L}$ ) of the Kerr rotation, which is dominant, the
QMOKE signal, which is most probably due to the second order spin-orbit
coupling {[}26{]}, is present. The derived ($M_{L}^{2}-M_{T}^{2}$
) and $M_{L}M_{T}$ angular variations show the behaviour expected
from the above equation. 

\begin{figure*}
\includegraphics[bb=20bp 200bp 550bp 595bp,clip,width=15cm]{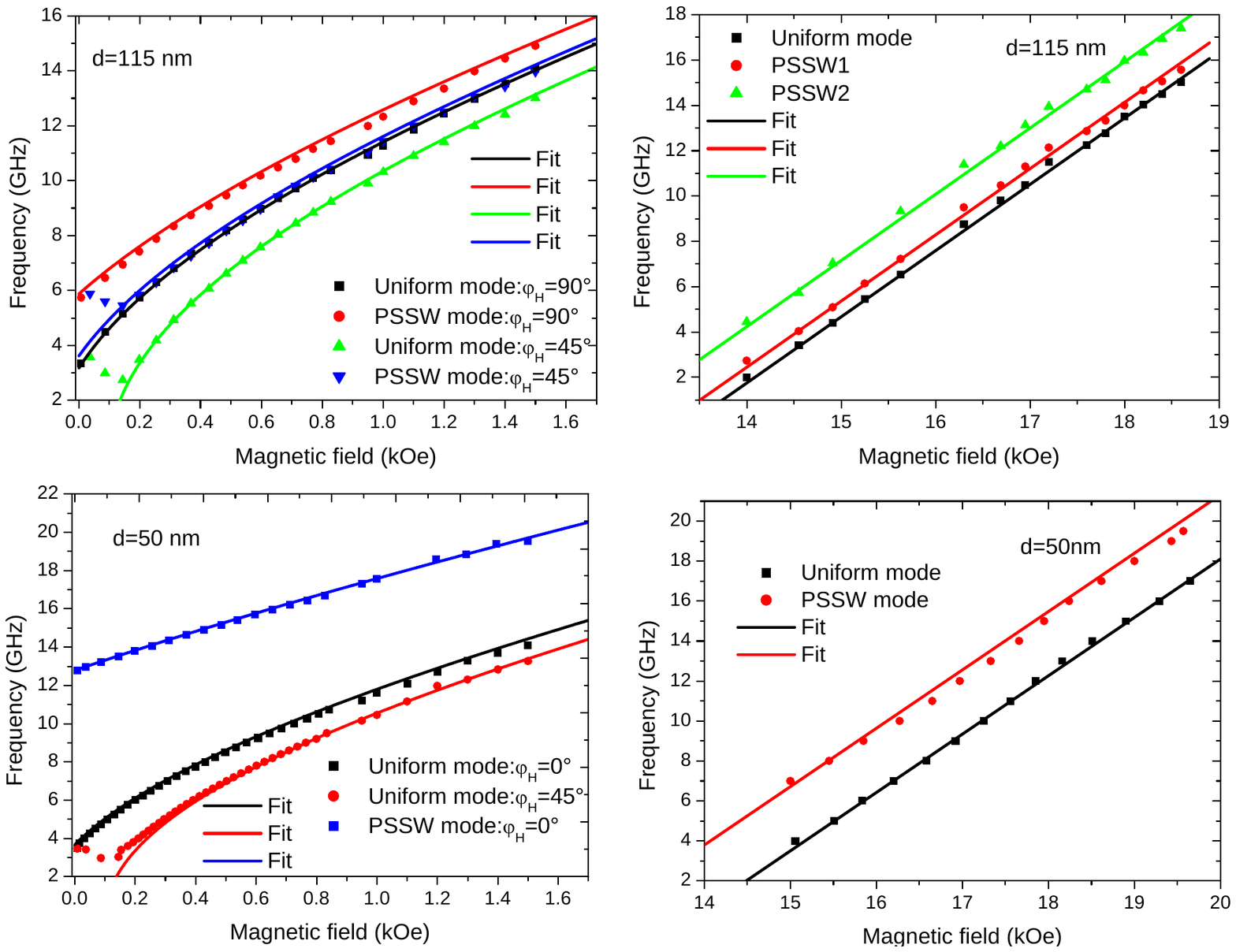}
\caption{(Colour online) Field dependence of the resonance frequency of the
uniform precession and of the two first perpendicular standing spin
wave excited (PSSW) mode of 115 nm Cr-capped and 50 nm Ta-capped Co$_{2}$FeAl
films. The magnetic field is applied perpendicular or in the film
plane. The fits are obtained using equations (6) and (7) with the
parameters indicated in the Table I.}
\end{figure*}

\begin{figure}
\includegraphics[bb=20bp 360bp 330bp 595bp,clip,width=8.5cm]{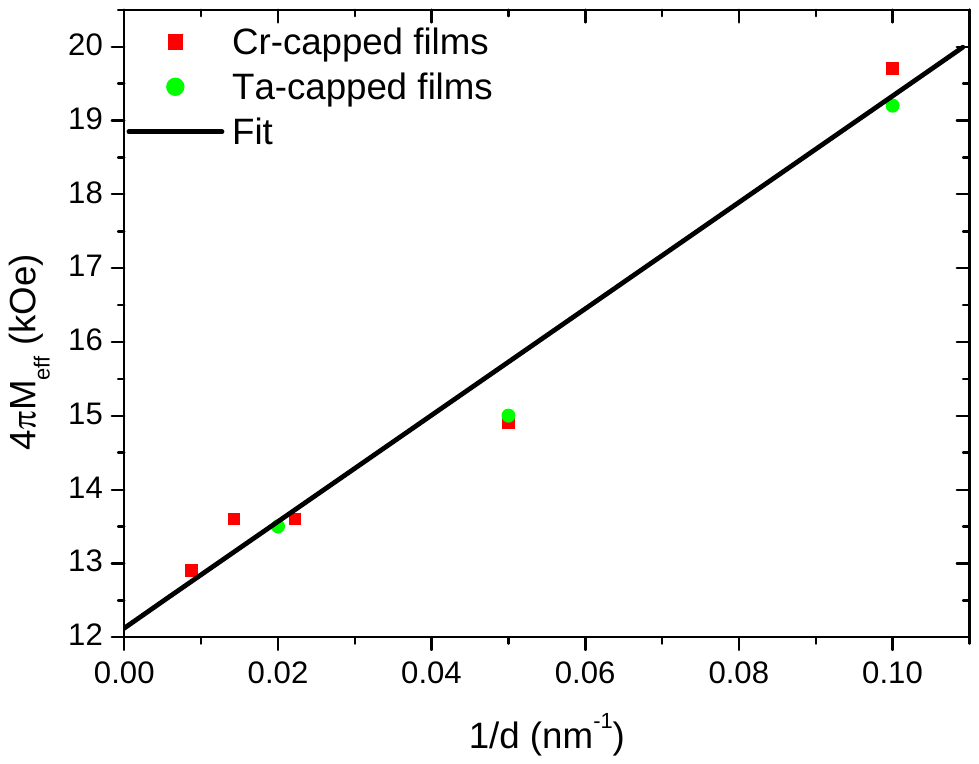}
\caption{(Colour online) Thickness dependence of the effective magnetization
($4\pi M_{eff}$) extracted from the fit of FMR measurements. The
solid lines are the linear fits.}
\end{figure}

The values for the amplitudes of the $2M_{L}M_{T}$ and of the ($M_{L}^{2}-M_{T}^{2}$
) contributions are the same within the experimental error for each
sample suggesting that the applied cubic model is correct. The offset
of the $M_{L}M_{T}$ contribution is smaller than the amplitudes,
but generally it follows the same trend as the amplitudes. As the
thickness decreases the amplitudes and the offset decrease, suggesting
that the chemical order progressively changes from the B2 to the A2
phase, as discussed above. Moreover, the amplitudes and offset values
of CFA are comparable to those measured for Co$_{2}$MnSi, which presents
the L2$_{1}$ phase {[}28{]}. The TBIIST results are discussed in
the following section, in order to allow for a comparison with the
data derived from the FMR study of the dynamic properties. 

\begin{figure*}
\includegraphics[bb=20bp 150bp 540bp 595bp,clip,width=15cm]{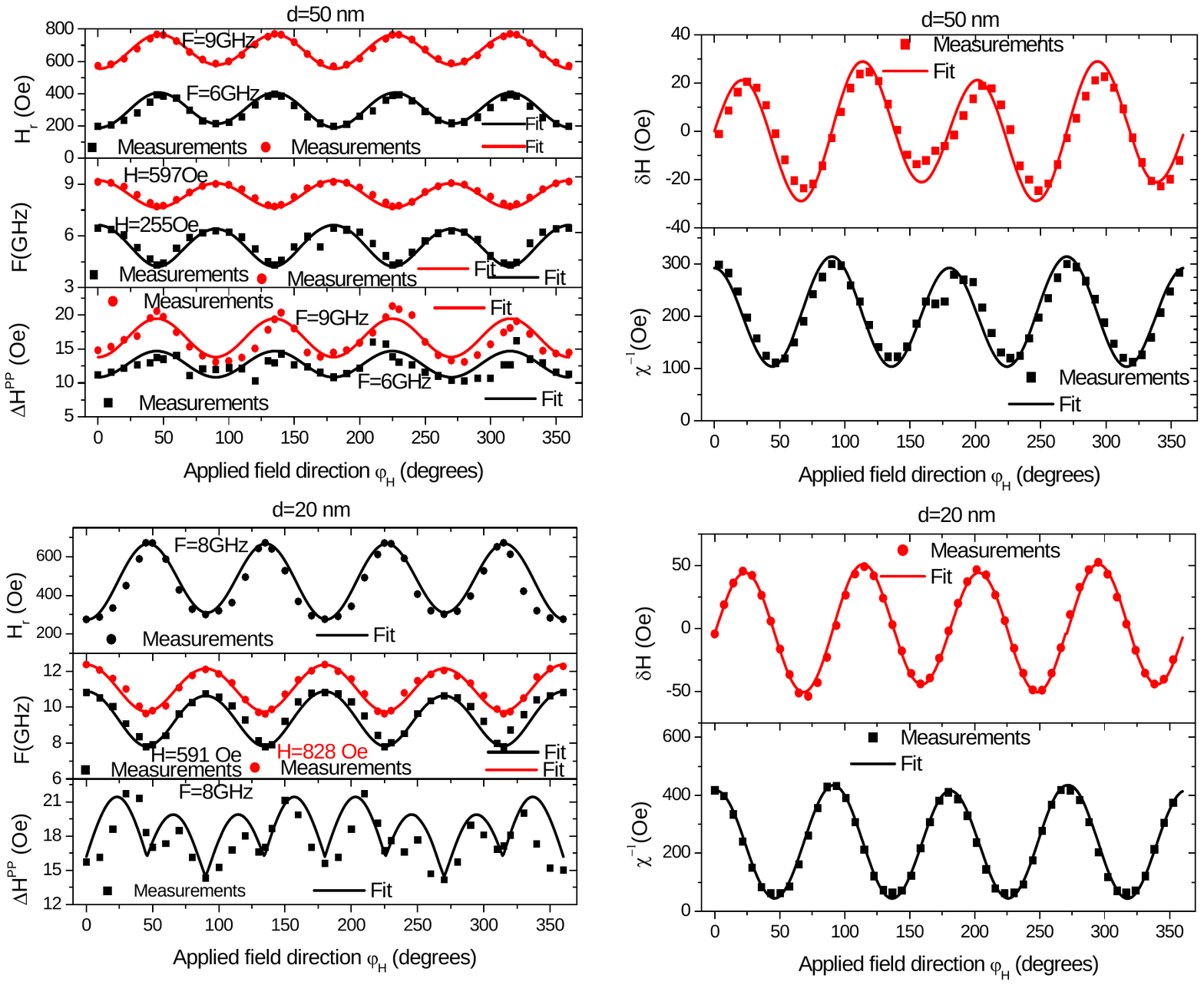}
\caption{(Colour online) Angular dependence of the resonance frequency ($F_{r}$),
resonance field ($H_{r}$), peak to peak field FMR linewidth ($\Delta H^{PP}$),
inverse susceptibility ($\chi^{-1}$) and the field offset ($\delta_{H}$)
of 50 nm and 20 nm thick Co$_{2}$FeAl Ta-capped thin films. The TBIIST
measurements were obtained using transverse static bias field $H_{B}=200$
Oe and 225 Oe respectively for 50 nm and 20 nm thick Co$_{2}$FeAl
films. The solid lines refer to the fit suing the above mentioned
models.}
\end{figure*}

\subsection{Dynamic properties}

\subsubsection{Exchange stiffness and effective magnetization}

The uniform precession and the first PSSW modes have been observed
in perpendicular and in-plane applied field configurations for samples
thicknesses down to 50 nm. For the thickest film (115 nm) it was even
possible to observe the second PSSW. For lower sample thickness, the
PSSW modes are not detected due their high frequencies over-passing
the available bandwidth (0-24 GHz). Typical in-plane and perpendicular
field dependences of the resonance frequencies of the uniform and
PSSW modes are shown on figure 7 for the 115 nm Cr- and the 50 nm
Ta-capped films. By fitting the data in figure 7 to the above presented
model, the gyromagnetic factor ($\gamma$), the exchange stiffness
constant ($A_{ex}$) and the effective magnetization $(4\pi M_{eff}$)
are extracted. The fitted $\gamma$ and $A_{ex}$ values are 2.92
GHz/kOe and 1.5 $\lyxmathsym{\textmu}$erg/cm, respectively: they
do not depend of the studied sample. The derived exchange constant
is in good agreement with the reported one by Trudel et al. {[}7{]}.
$M_{eff}=H_{eff}/4\pi$

Figure 8 plots out the extracted effective magnetization $4\pi M_{eff}$
versus the film thickness $1/d$. It can be seen that $M_{eff}$ follows
a linear variation. This allows to derive the perpendicular surface
anisotropy coefficient $K_{\perp S}$: $K_{\perp S}=-1.8$ erg/cm$^{2}$.
The limit of $4\pi M_{eff}$ when $1/d$ tends to infinity is equal
to 12.2 kOe: within the above mentioned experimental precision about
the magnetization at saturation it does not differ from $4\pi M_{S}$.
We conclude that the perpendicular anisotropy field derives from a
surface energy term; being negative, it provides an out-of-plane contribution.
It may originate from the magneto-elastic coupling arising from the
interfacial stress due to the substrate.

\subsubsection{Magnetic anisotropy }

Figure 9 shows the angular dependences of the resonance field (at
fixed frequency) and of the resonance frequency (at fixed applied
field) compared to the static TBIIST measurements for three different
CFA films. Both FMR and TBIIST data show that the angular behavior
is governed by a superposition of uniaxial and fourfold anisotropy
terms with the above-mentioned easy axes. As noticed above, the symmetry
properties of the epitaxial observed films agree with the principal
directions of the fourfold contribution. The fourfold and uniaxial
anisotropy fields extracted from the fit of the experimental TBIIST
and FMR data using the above-presented model are drawn on figure 10
and summarized in Table I: the compared results issued from the two
techniques are in excellent agreement. For all the samples the fourfold
anisotropy is dominant. While the uniaxial anisotropy field ($H_{2}$)
of the Cr-capped films is small and does not seem to depend upon the
thickness, in the Ta-capped films $H_{2}$ is higher, maybe due to
interface effects, and is a decreasing function of the thickness (Figure
10). As suggested previously, we believe that the uniaxial anisotropy
is induced by the stepping of the substrates, probably resulting from
a small miscut along their $[100]$ crystallographic direction corresponding
to the $[110]$ studied films. The reduced effect of the steps of
the substrate when the thickness increases could then explain the
thickness dependence of $H_{2}$. However, up to now we have no completely
satisfying interpretation of the presence of $H_{2}$ and of its variations
versus the nature of the film capping. 

The fourfold anisotropy fields ($H_{4}$) are comparable for Cr- and
Ta-capped films and decrease when their thickness increases, as seen
in figure 10. For large values of $d$ (45nm or higher) $H_{4}$ lies
around 200 Oe and shows a small linear variation versus the in-plane
strain $\varepsilon_{\parallel}$, as shown in the insert of figure
10. This evolution confirms a direct correlation between the H4 field
and the in-plane biaxial strain for the films with thicknesses above
45 nm. At smaller values of $d$ (10 or 20 nm) a large increase of
$H_{4}$, up to 920 Oe, is observed. It is presumably related to the
B2$\Rightarrow$A2 phase transition observed through X-rays diffraction.
The observed symmetry argues for a magneto-crystalline contribution,
which, as previously observed {[}29, 30{]}, would be higher in phase
A2 than in phase B2.

\subsubsection{FMR linewidth}

\begin{figure}
\includegraphics[bb=20bp 380bp 290bp 595bp,clip,width=8.5cm]{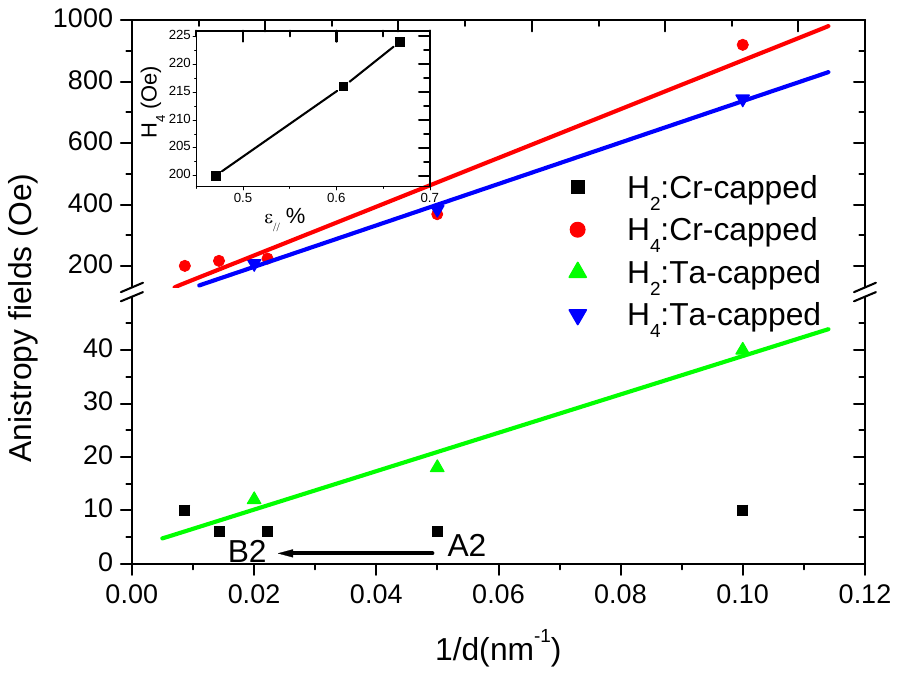}
\caption{(Colour online) Thickness dependence of the uniaxial ($H_{2}$) and
the fourfold anisotropy fields ($H_{4}$) extracted from the fit of
FMR measurements. The solid lines are the linear fits. The inset shows
the evolution of the $H_{4}$ field, for the 45, 70 and 115 nm thick
samples, with the in-plane biaxial strain.}
\end{figure}

In figure 9, the FMR peak to peak linewidth (($\Delta H^{PP}$) is
plotted as a function of the field angle $\varphi_{H}$ for the 50
nm and 20 nm Ta-capped CFA films using three driving frequencies:
6, 8, and 9 GHz.$\Delta H^{PP}$ is defined as the field difference
between the extrema of the sweep-field measured FMR spectra. All the
other samples show a qualitatively similar behaviour to one of the
samples presented here. The positions of the extrema depend on the
sample. The observed pronounced anisotropy of the linewidth cannot
be due to the Gilbert damping contribution, which is expected to be
isotropic, and must be due to additional extrinsic damping mechanisms.
In the 50 nm thick sample, the $\Delta H^{PP}$ angular variation
shows a perfect fourfold symmetry (in agreement with the variation
of the resonance position). Such behaviour is characteristic of two
magnon scattering. This effect is correlated to the presence of defects
preferentially oriented along specific crystallographic directions,
thus leading to an asymmetry (see equation (11)). Concerning the 20
nm thick film, the in-plane angular dependence of $\Delta H^{PP}$
is less simple and shows eight maxima, that is expected from a mosaicity
driven linewidth broadening. It can be seen that a smaller fourfold
symmetry (four maxima) is superimposed on the eight maxima, indicating
that two magnon scattering is still present. Therefore, the entire
angular dependence of the FMR linewidth in our samples can be explained
as resulting of the four contributions appearing in equation (8). 

\begin{figure}
\includegraphics[bb=20bp 370bp 330bp 595bp,clip,width=8.5cm]{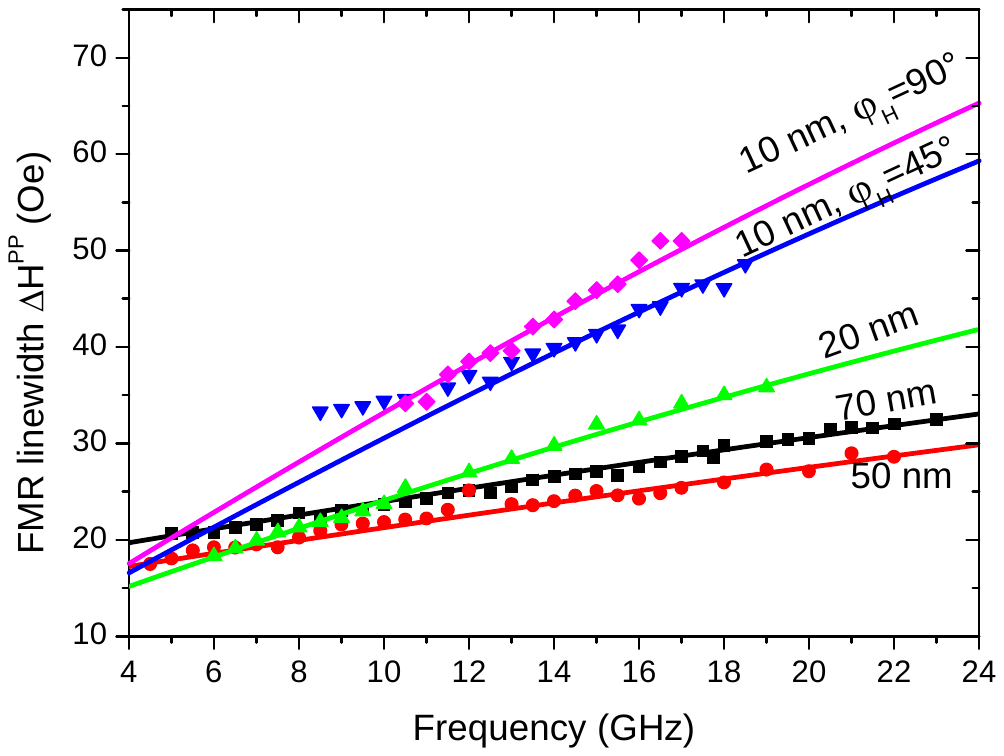}
\caption{(Colour online) Frequency dependence of the easy axis ($\varphi_{H}=0$)
peak to peak field FMR linewidth ($\Delta H^{PP}$) for Co$_{2}$FeAl
thin. The solid lines refer to the fit using equations (8-11).}
\end{figure}

In figure 11, $\Delta H^{PP}$ for the field parallel to an easy axis
and a hard axis ($\varphi_{H}=45\lyxmathsym{\textdegree}$ for 10
nm thick sample) of the fourfold anisotropy is plotted as a function
of driving frequency for all samples. An apparently extrinsic contribution
to linewidth was observed, which increased with decreasing film thickness.
It should be mentioned that the observed linear increase of the linewidth
with frequency in figure 11 maybe due to Gilbert damping but other
relaxation mechanisms can lead to such linear behaviour. Therefore,
only an effective damping parameter $\alpha_{eff}$ can be extracted
from the slope of the curves and ranges between about 0.00154 for
the easy axis of the 50 nm thick film and 0.0068 for easy axis the
thinnest film. The pertinent parameters could thus be, in principle
derived from the conjointly analysis of the frequency and angular
dependence of $\Delta H^{PP}$. However, due to the limited experimental
precision, some additional hypotheses are necessary in order to allow
for a complete determination of the whole set of parameters describing
the intrinsic Gilbert damping and the two magnon damping. A detailed
analysis is presented in the appendix. Using the previously reported
value: $\alpha=1.1\times10^{-3}$ {[}31{]}, which is in agreement
with our experimental results, we were able to $\Gamma_{0}$ for each
film. $\Gamma_{0}$, $\Gamma_{2}$,$\Gamma_{4}$, $\varphi_{2}$,
$\varphi_{4}$ are listed in Table II which also contains the parameters
describing the damping effects of the mosaïcity ($\Delta\varphi_{H}$)
and of the inhomogeneity contribution ($\Delta H^{inh}$).

The two magnon and the mosaïcity ($\Delta\varphi_{H}$) contributions
to $\Delta H^{PP}$ increase when the thickness decreases, probably
due to the progressive above reported loss of chemical order. The
increase of the residual inhomogeneities linewidth ($\Delta H^{inh}$)
with the thickness is most probably due the increase of defects and
roughness. The uniaxial term $\Gamma_{2}$ is observed only in the
thinnest (20 and 10 nm) samples. As expected, $\varphi_{4}=0$, but
the sign of $\Gamma_{4}$ is sample dependent. Finally, it is important
to notice that the very low value of the intrinsic damping in the
studied samples allows for investigating the extrinsic contributions.

\section{Conclusion}

Co2FeAl films of various thicknesses (10 nm$\leq d\leq115$ nm)) were
prepared by sputtering on a (001) MgO substrate. They show full epitaxial
growth with chemical order changing from B2 to A2 phase as thickness
decreases. MOKE and VSM hysteresis loops obtained with different field
orientations revealed that, depending on the direction of the in-plane
applied field, two or one jump switching occur, due to the superposition
of uniaxial and fourfold anisotropies. The samples present a quadratic
MOKE contribution with decreasing amplitudes as the CFA thickness
decreases. The microstrip ferromagnetic resonance (MS-FMR) and the
transverse biased initial inverse susceptibility and torque (TBIIST)
methods have been used to study the dynamic properties and the anisotropy.
The in-plane anisotropy presents two contributions, showing a fourfold
and a twofold axial symmetry, respectively. A good agreement concerning
the relevant in-plane anisotropy parameters deduced from the fit of
MS-FMR and TBIIST measurements has been obtained. The fourfold in-plane
field shows a thickness dependence behavior correlated to the thickness
dependence of the chemical order and strain in samples. The angular
and frequency dependences of the FMR linewidth are governed by two
magnon scattering, mosaïcity and by a sample independent Gilbert damping
equal to 0.0011

\section*{Appendix}

In the section dealing with the discussion of the FMR linewidth measurements
we stated that the conjointly analysis of the frequency and angular
dependence of $\Delta H^{PP}$ does not allow for the determination
of all the parameters given in equation (8) and additional hypothesis
should be done. The aim of this appendix is to clarify the manner
in which the parameters summarized in Table II is done.

For most of the exploitable measurements the microwave frequency f
during the $\Delta H^{PP}$ measurements is not larger than $f_{0}$
and generally smaller ($f_{0}$ varies from 18.5 to 28.5 GHz, depending
on the film thickness). It then results that the two magnon damping
is practically proportional to f and that the sum of the Gilbert and
of the two magnon damping terms reads as (see equations (9) and (11)):

\begin{multline}
\Delta H^{Gi+2mag}\cong((\frac{\alpha}{\sqrt{3}}+\frac{\Gamma_{0}}{2H_{eff}})+\frac{\Gamma_{2}}{2H_{eff}}\cos2(\varphi_{H}-\varphi_{2})\\
+\frac{\Gamma_{4}}{2H_{eff}}\cos4(\varphi_{H}-\varphi_{4}))\frac{4\pi}{\gamma}f
\end{multline}

It is not possible to completely identify the respective contributions
of the Gilbert and of the two magnon damping, only according to equation
(13). The quasi-linear variation versus the frequency (Fig. 11) observed
for $\Delta H^{PP}$ allows for defining an effective damping parameter
$\alpha_{eff}$, which, is angle dependent due to two magnon scattering.
The experimentally derived coefficient $\alpha_{eff}$, from the linear
fit of data presented in figure 11, varies from 0.0068 to 0.00154.
Furthermore, the measured angular variation of the linewidth allows
for evaluating ($\Gamma_{2}$, $\varphi_{2}$) and ($\Gamma_{4}$,
$\varphi_{4}$) but, concerning the isotropic terms appearing in equation
(13), only the sum $\alpha+\frac{\sqrt{3}\Gamma_{0}}{2H_{eff}}$ can
be derived. However, remembering that $\alpha$ cannot be negative,
the maximal available value of $\Gamma_{0}$ (corresponding to $\alpha=0$
) is easily found. Moreover, a lowest value can be obtained for $\Gamma_{0}$
noticing that equation (13) can also be written:

\begin{multline}
\Delta H^{Gi+2mag}\cong((\frac{\alpha}{\sqrt{3}}+\frac{\Gamma_{0}-\left|\Gamma_{2}\right|-\left|\Gamma_{4}\right|}{2H_{eff}})+\\
\frac{\left|\Gamma_{2}\right|}{2H_{eff}}(1\pm\cos2(\varphi_{H}-\varphi_{2}))+\\
\frac{\left|\Gamma_{4}\right|}{2H_{eff}}(1\pm\cos4(\varphi_{H}-\varphi_{4})))\frac{4\pi}{\gamma}f
\end{multline}

where the adequate third and the fourth terms represent the twofold
and the fourfold contributions, which take into account that both
of them are necessarily non-negative for any value of $\varphi_{H}$.
The additional residual two magnon isotropic contribution cannot be
negative. Hence: $\Gamma_{0}>\left|\Gamma_{2}\right|+\left|\Gamma_{4}\right|$.

Introducing this minimal accessible value of $\Gamma_{0}$ , ($\left|\Gamma_{2}\right|+\left|\Gamma_{4}\right|$
), the maximal value of the Gilbert coefficient $\alpha$ is then
easily obtained. To summarize, for each sample the experimental data
provide the allowed intervals for $\alpha$ and for $\Gamma_{0}$,
respectively {[}0, $\alpha_{min}${]} and {[}($\left|\Gamma_{2}\right|+\left|\Gamma_{4}\right|$),
$\Gamma_{0Max}${]}, and indeed, the chosen value of $\alpha$ within
{[}0, $\alpha_{min}${]} allows for deducing $\Gamma_{0}$ . The smallest
calculated interval for $\alpha$, equal to {[}0, $1.4\times10^{-3}${]}
is obtained for the 70 nm film. A previous publication by Mizukami
el al. {[}31{]} has concluded to a Gilbert coefficient equal to: $\alpha=1.1\times10^{-3}$.
We then stated that $\alpha=1.1\times10^{-3}$ and, consequently,
we were able to deduce $\Gamma_{0}$ for each film. $\Gamma_{0}$,
$\Gamma_{2}$,$\Gamma_{4}$, $\varphi_{2}$, $\varphi_{4}$ are listed
in Table II which also contains the parameters describing the damping
effects of the mosaïcity and of the inhomogeneity.


\begin{thebibliography}{References}
\bibitem{key-1}{[}1{]} A. Yanase and K. Siratori, J. Phys. Soc. Jpn.
53, 312 (1984) 

\bibitem{key-30}{[}2{]} Z. Zhang and S. Satpathy, Phys. Rev. B 44,
13319 (1991)

\bibitem{key-29}{[}3{]} K. Schwarz, J. Phys. F: Met. Phys. 16, L211
(1986) 

\bibitem{key-28}{[}4{]} J. H. Park, E. Vescovo, H. J. Kim, C. Kwon,
R. Ramesh, and T. Venkatesan, Nature (London) 392, 794 (1998)

\bibitem{key-27}{[}5{]} H. C. Kandpal, G. H. Fecher, and C. Felser,
J. Phys. D 40, 1507 (2007) 

\bibitem{key-26}{[}6{]} R. A. de Groot, F. M. Mueller, P. G. van
Engen and K. H. J. Buschow, Phys. Rev. Lett. 50, 2024 (1983)

\bibitem{key-25}{[}7{]} S. Trudel, O. Gaier, J. Hamrle, and B. Hillebrands,
J. Phys. D 43, 193001 (2010)

\bibitem{key-24}{[}8{]} W. H. Wang, E. Liu, M. Kodzuka, H. Sukegawa,
M. Wojcik, E. Jedryka, G. H. Wu, K. Inomata, S. Mitani, and K. Hono,
Phys. Rev. B 81, 140402 (R) (2010)

\bibitem{key-23}{[}9{]} W. H. Wang, H. Sukegawa, and K. Inomata,
Phys. Rev. B 82, 092402 (2010)

\bibitem{key-22}{[}10{]} D. Berling, S. Zabrocki, R. Stephan, G.
Garreau, J.L. Bubendorff, A. Mehdaoui, D. Bolmont, P. Wetzel, C. Pirri
and G. Gewinner, J. Magn. Magn. Mat. 297, 118 (2006) 

\bibitem{key-21}{[}11{]} M. S. Gabor, T. Petrisor Jr., and C. Tiusan,
M. Hehn and T. Petrisor, Phys. Rev. B 84, 134413 (2011) 

\bibitem{key-20}{[}12{]} D. Fruchart, R. Fruchart, Ph. L\textquoteright{}Héritier,
K. Kanematsu, R. Madar, S. Misawa, Y. Nakamura, P. J. Ziebeck, and
K. R. A. Webster, in Magnetic Properties of Metals, edited by H. P.
J. Wijn, Landolt- Börnstein, New Series, Group III, Vol. 19c, Pt.
2 (Springer-Verlag, Berlin, 1986).

\bibitem{key-19}{[}13{]} K. Inomata, N. Ikeda, N. Tezuk, R. Goto,
S. Sugimoto, M. Wojcik and E. Jedryka, Sci. Technol. Adv. Mater. 9
(2008) 014101 

\bibitem{key-18}{[}14{]} M. Belmeguenai, F. Zighem, Y. Roussigné,
S-M. Chérif, P. Moch, K. Westerholt, G. Woltersdorf, and G. Bayreuther
Phys. Rev. B 79, 024419 (2009)

\bibitem{key-17}{[}15{]} M. Belmeguenai, F. Zighem, T. Chauveau,
D. Faurie, Y. Roussigné, S-M. Chérif, P. Moch, K.Westerholt and P.
Monod, J. Appl. Phys. 108, 063926 (2010)

\bibitem{key-16}{[}16{]} Kh. Zakeri, J. Lindner, I. Barsukov, R.
Meckenstock, M. Farle, U. von Hörsten, H. Wende, W. Keune, J. Rocker,
S. S. Kalarickal, K. Lenz, W. Kuch, and K. Baberschke, Phys. Rev.
B 76, 104416 (2007)

\bibitem{key-15}{[}17{]} H. Lee, Y.-H. A. Wang, C. K. A. Mewes, W.
H. Butler, T. Mewes, S. Maat, B. York, M. J. Carey, and J. R. Childress,
Appl. Phys. Lett. 95, 082502 (2009)

\bibitem{key-14}{[}18{]} H. Kurebayashi, T. D. Skinner, K. Khazen,
K. Olejnik, D. Fang, C. Ciccarelli, R. P. Campion, B. L. Gallagher,
L. Fleet, A. Hirohata, and A. J. Ferguson, Appl. Phys. Lett 102, 062415
(2013)

\bibitem{key-13}{[}19{]} K D Sossmeier, F Beck, R C Gomes, L F Schelp
and M Carara, J. Phys. D: Appl. Phys. 43, 055003 (2010)

\bibitem{key-12}{[}20{]} Y. Y. Zhou, X. Liu, J. K. Furdyna, M. A.
Scarpulla and O. D. Dubon, Phys. Rev. B 80, 224403 (2009) 

\bibitem{key-11}{[}21{]} W. Platow, A. N. Anisimov, G. L. Dunifer,
M. Farle, and K. Baberschke, Phys. Rev. B 58, 5611 (1998)

\bibitem{key-10}{[}22{]} R. Arias and D. L. Mills, Phys. Rev. B 60,
7395 (1999)

\bibitem{key-8}{[}23{]} D. L. Mills and R. Arias, Physica B 384,
147 (2006)

\bibitem{key-9}{[}24{]} R. Arias and D. L. Mills, J. Appl. Phys.
87, 5455 (2000).

\bibitem{key-7}{[}25{]} R. P. Cowburn, S. J. Gray, J. Ferré, J. A.
C. Bland, and J. Miltat, J. Appl. Phys. 78, 7210 (1995).

\bibitem{key-6}{[}26{]} K. Postava, D. Hrabovsky, J. Pistora, A.
R. Fert, S. Visnovsky, and T. Yamaguchi, J. Appl. Phys. 91, 7293 (2002).

\bibitem{key-5}{[}27{]} R.M. Osgood III, S.D. Bader, B. M. Clemens,
R.L. White, H. Matsuyama, J. Magn. Magn. Mater. 182, 297 (1998). 

\bibitem{key-4}{[}28{]} G. Wolf, J. Hamrle, S. Trudel, T. Kubota,
Y. Ando and B. Hillebrands, J. Appl. Phys.110, 043904 (2011) 

\bibitem{key-3}{[}29{]} O. Gaier, J. Hamrle, S. J. Hermsdoerfer,
H. Schultheiß, B. Hillebrands, Y. Sakuraba, M. Oogane, and Y. Ando,
J. Appl. Phys. 103, 103910 (2008)

\bibitem{key-2}{[}30{]} S.Trudel, G. Wolf, J. Hamrle, B. Hillebrands,
P. Klaer, M. Kallmayer, H.-J. Elmers, H. Sukegawa, W. Wang, and K.
Inomata, Phys. Rev. B 83, 104412 (2011) 

\bibitem{key-1}{[}31{]} S. Mizukami, D. Watanabe, M. Oogane, Y. Ando,
Y. Miura, M. Shirai and T. Miyazaki, J. Appl. Phys.105, 07D306 (2009) \end{thebibliography}
\end{document}